\documentclass{Rinton-P9x6}

\def\nue{\nu_{e}}
\def\num{\nu_{\mu}}
\def\nut{\nu_{\tau}}
\def\nmnt{$\nu_{\mu}\leftrightarrow\nu_{\tau}$~}
\def\nenm{$\nu_{e}\leftrightarrow\nu_{\mu}$~}
\def\lsim{\lower.7ex\hbox{${\buildrel < \over \sim}$}}
\def\gsim{\lower.7ex\hbox{${\buildrel > \over \sim}$}}

\begin{document}

\title{Current status of the K2K experiment\footnote{Talk
at Cairo International Conference on High Energy Physics (CICHEP2001),
Cairo, Egypt, January~9-14,~2001.}}

\author{Yuichi Oyama\\
for K2K collaboration}

\address{Institute of Particles and Nuclear Studies,\\
High Energy Accelerator Research Organization (KEK)\\
Oho 1-1, Tsukuba, Ibaraki 305-0801, Japan\\
E-mail:~~yuichi.oyama@kek.jp\\}

\maketitle

\abstracts{
Current status of the K2K (KEK to Kamioka) long-baseline
neutrino-oscillation experiment is presented.
}

\section{Introduction}
The K2K experiment\cite{Nishikawa,detect,cpvio,moriond}
is the first long-baseline neutrino-oscillation experiment
with hundreds of km distance
using an accelerator-based neutrino beam.
The nominal sensitive region in the neutrino-oscillation parameters
is $\Delta m^{2} > 3\times 10^{-3}$eV$^{2}$.
This covers the parameter region suggested by the atmospheric neutrino
anomaly observed by several underground experiments,\cite{Kamioka,IMB,Soudan}
and confirmed by Super-Kamiokande(SK).\cite{SuperK}

An overview of the K2K experiment is as follows.
Almost a pure wide-band $\nu_{\mu}$ beam from $\pi^{+}$ decays is generated
in the KEK 12-GeV/c Proton Synchrotron (PS) and a neutrino beam-line,
and is detected in SK at a distance of 250km.
Various beam monitors along the beam line and two different types of
front detectors (FDs) are also constructed at the KEK site.
The FDs are a 1kt water Cherenkov detector~(1KT), which is
a miniature of the SK detector, and a so-called fine-grained detector (FGD),
which is composed of a scintillating fiber tracker~(SFT)\cite{scifi},
trigger counters~(TRG), lead glass counters~(LG)
and a muon range detector~(MRD).
Since the design and performance of these components
as well as the properties of the neutrino beam
were already described precisely
in previous articles\cite{cpvio,moriond}, they are not discussed here.

The K2K experiment was successfully started in early 1999, and data were
recorded in January to March, and May to June in 2000.
The total data-taking period in 1999 and 2000 was 112.2~days.
The accumulated beam intensity was
$22.9 \times 10^{18}$ protons on target (p.o.t.),
which is about 20\% of the goal of the experiment, $10^{20}$~p.o.t.

\section{Study of Neutrino beam properties in KEK site}
The characteristics of the neutrino beam in the KEK site
were examined using FDs and beam monitors.
In this section, the present status of analyses on (1)the neutrino beam direction,
(2)the neutrino beam intensity and its stability, (3)the $\nue/\num$ ratio,
and (4)the neutrino energy spectrum are presented.

\subsection{Neutrino beam direction}
The neutrino beam-line was constructed with a GPS position survey\cite{GPS},
and the alignment of the beam-line, FDs and SK is better than 0.1~mrad.
The neutrino beam direction relative to the beam-line was measured
with the muon monitors and MRD independently. 

\begin{figure}[b!]
\centerline{
\epsfig{file=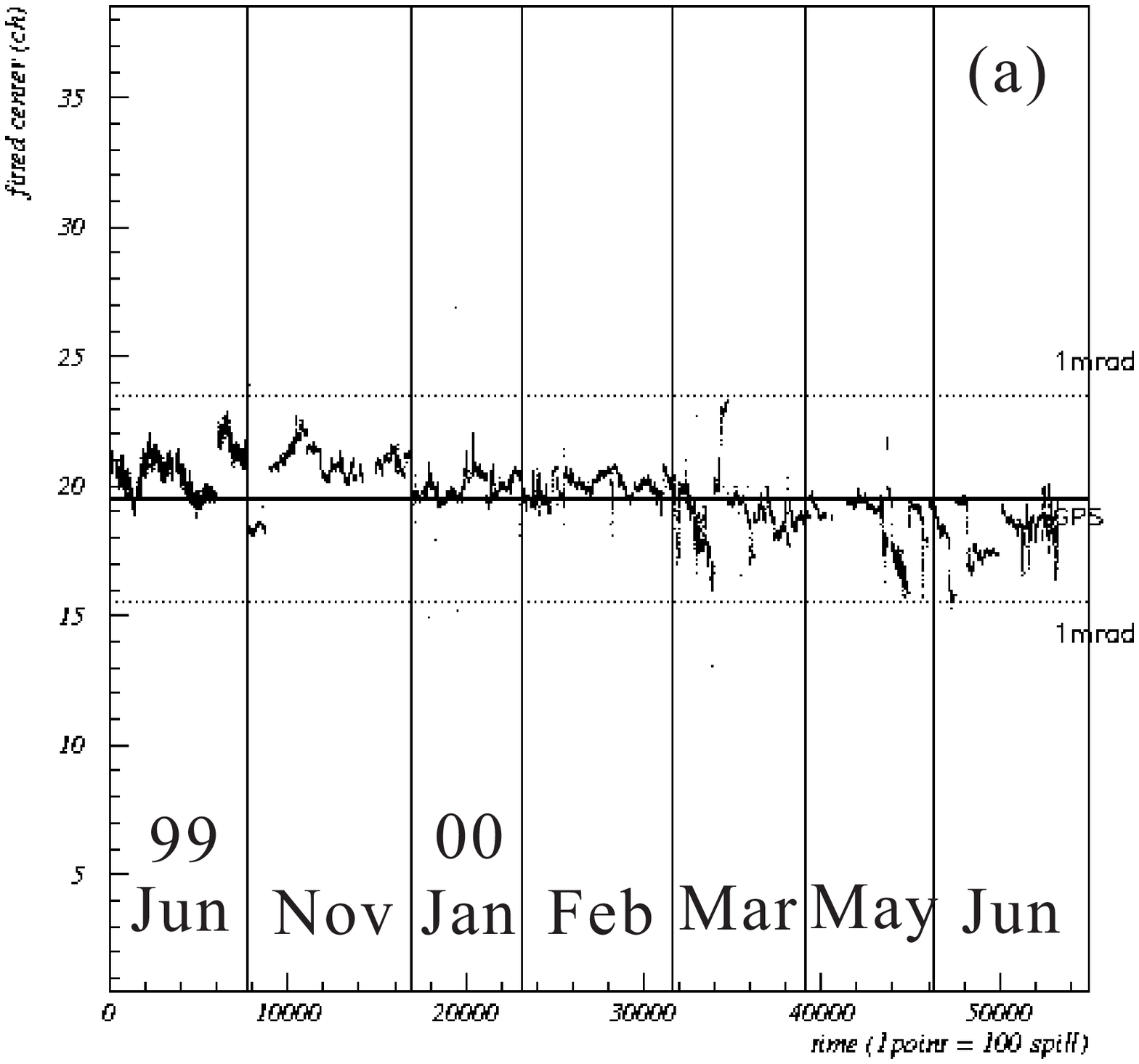,height=5.cm}
\hskip 1.1cm
\epsfig{file=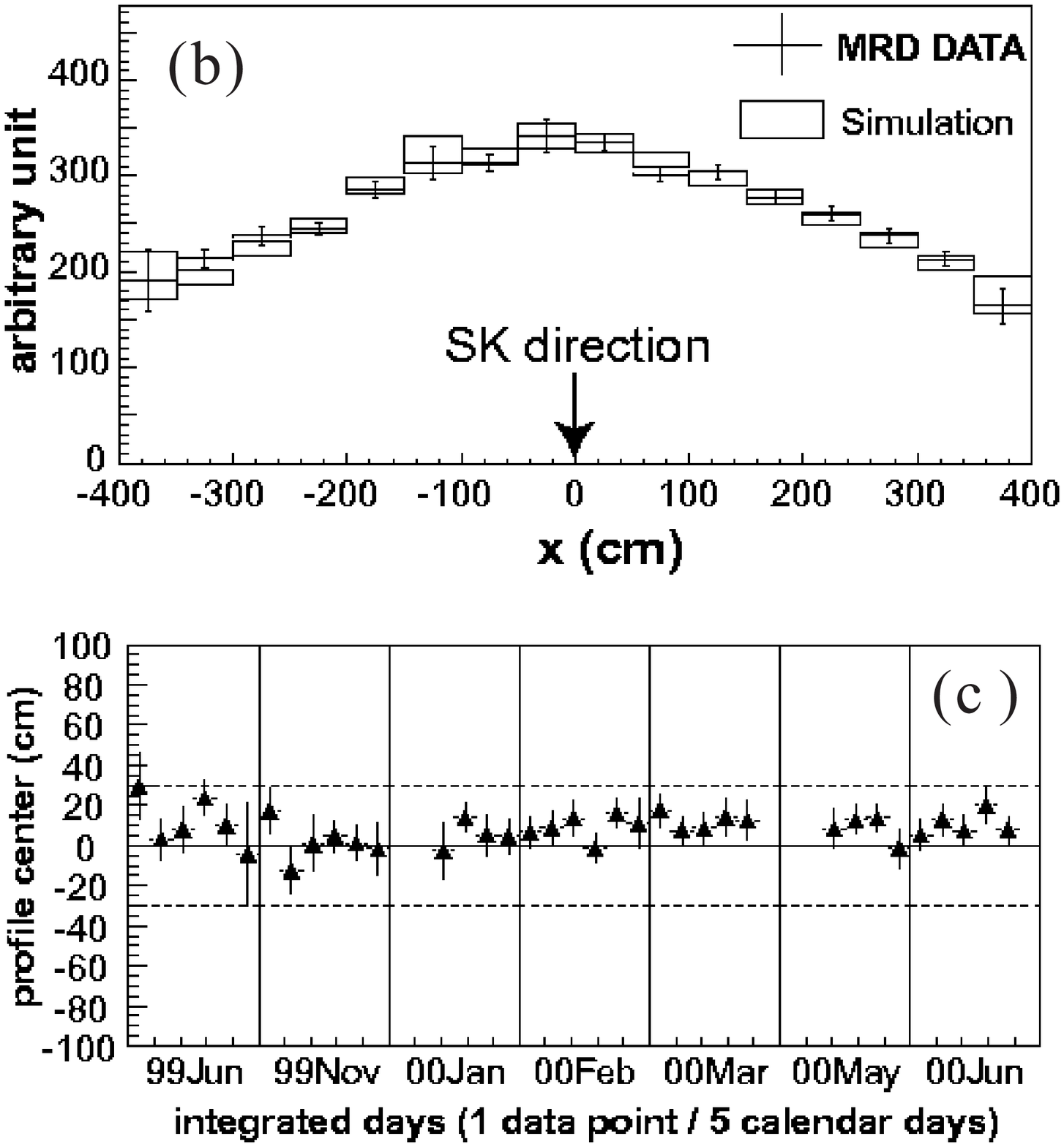,height=5.cm}
}
\caption{
Examination of the neutrino beam direction.
They are (a)~time variation of the beam direction from the muon monitors,
(b)~vertex distribution of neutrino interactions in MRD, and
(c)~time variation of the profile center. In (a) and (c),
the SK direction and 1~mrad off axis are shown by the solid and dashed lines,
respectively.
}
\label{fig1}
\end{figure}

The muon monitor consists of a segmented ionization
chamber and an array of silicon pad detectors, which are located
downstream of the beam dump. Their position resolution
is about 2cm, corresponding to an angular resolution of 0.1mrad.
Because $\num$ and muons originate in the same pion decay in the decay volume,
the $\num$ beam direction can be examined from the profile
center of the muon beam.
The time variation of the profile center is plotted in Figure~\ref{fig1}(a).
The direction of the muon beam agrees with the beam-line within 1~mrad.

The neutrino beam direction is also measured using neutrino interactions
in MRD. The distribution of the vertex position is plotted in Figure~\ref{fig1}(b).
The center of the beam profile agrees with the SK direction within 1~mrad.
The time variation of the beam center, also plotted in Figure~\ref{fig1}(c),
shows that the steering of the beam direction is stable,
and is consistent with the results from the muon monitors.

The energy spectrum of the neutrino beam
is expected to be uniform within 3~mrad from the center of
the beam axis.
On the other hand, the angular acceptance of the SK detector
from the KEK site is about 0.2~mrad.
Therefore, the adjustment of the neutrino beam
direction, ($<$ 1~mrad), is sufficient.

\begin{figure}[b!]
\centerline{
\epsfig{file=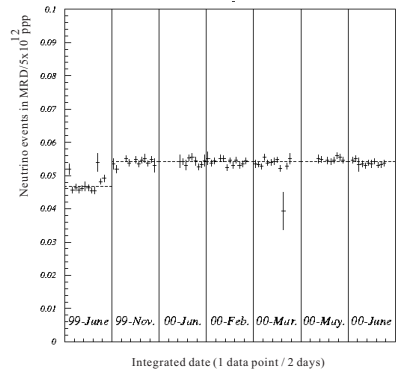,height=5.cm}
}
\caption{
Time variation of the neutrino event numbers in MRD.
The denominator of the vertical axis, $5\times 10^{12}$ppp
(protons per pulse), is a nominal beam intensity in one spill.
The June 99 data is smaller than the other periods because
the current of the magnetic horn was different.
The neutrino beam intensity is stable within
the statistical errors.
}
\label{fig2}
\end{figure}

\subsection{Neutrino beam intensity and its stability}

The neutrino beam intensity can be estimated from
absolute numbers of the neutrino interactions in 1KT, SFT and MRD,
and a comparison with Monte-Carlo expectations.

The Monte-Carlo simulation is based on GEANT\cite{GEANT}
with a detailed description of 
the materials and magnetic fields in the target region and the decay volume.
It uses as input a measurement of the primary-beam intensity
and profile at the target.
Primary proton interactions on aluminum are modeled with
a parameterization of hadron production
data\cite{hadr}. Other hadronic interactions are treated by 
GEANT-CALOR\cite{GCALOR}.

The number of neutrino interactions in 1KT, SFT and MRD
are consistent with each other, and
agree with the Monte-Carlo expectations within the systematic errors.

The stability of the beam intensity is contaneously measured from
the neutrino event rate in MRD because of its large statistics.
The time variation of the neutrino interactions in MRD
is shown in Figure~2.
The neutrino beam intensity is stable within a few \%.
Although the statistics is poor,
the neutrino event rates in SFT and 1KT are also stable.

\subsection{$\nue/\num$ ratio}
The $\nue/\num$ ratio of the neutrino beam at the KEK site
was measured by 1KT and FGD.
The idea of the measurements is given in Ref.4.
An analysis with 1KT is still under way, and no numerical result has been
obtained yet.
On the other hand, a very preliminary result with 
FGD is reported to be
(1.8$\pm$0.6\hbox{${{+0.8}\atop{-1.0}}$})\%,\cite{Yoshida}
where the expectation based on a Monte-Carlo simulation is 1.3\%.
The original neutrino beam has been proved to be almost
a pure $\num$ beam.

\begin{figure}[b!]
\centerline{
\psfig{file=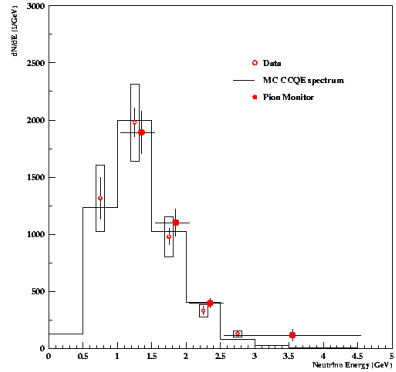,height=5.cm}
}
\caption{
Neutrino energy spectrum in FGD. They are ($\circ$)~measurement from
the quasi-elastic interactions in SFT, ($\bullet$)~calculation
based on a pion monitor measurement, ($-$)~Monte-Carlo simulation.
}
\label{fig3}
\end{figure}

\subsection{Neutrino energy spectrum}

The neutrino energy spectrum was studied with FGD and the pion monitor,
and compared with the Monte Carlo expectation.

To determine the neutrino energy spectrum from
neutrino interactions in the FGD,
quasi-elastic interactions of muon neutrinos, $\nu_{\mu}N \rightarrow \mu N'$,
in SFT were employed.
This is because most of the neutrino energy is transfered to the muons in
quasi-elastic interactions and
the neutrino energy can be directly calculated from the energy
and travel direction of the secondly muons.
The muon energy distribution obtained from quasi-elastic
interactions in SFT is shown in Figure~\ref{fig3}

The pion monitor was a gas Cherenkov detector with a spherical mirror
and R-C318 gas. The kinematic distribution of the pion beam was
calculated from the intensity and shape of the Cherenkov light in the
focus plane.
The energy spectrum and profile of the neutrino beam
can be calculated from a simple kinematics of the pion decay.
The neutrino energy distribution in FGD calculated from the
pion monitor data is shown in Figure~\ref{fig3}.

The expected neutrino energy spectrum was also calculated by the
Monte Carlo simulation described above. The result is also plotted
in Figure~\ref{fig3}. The agreement of 3 distributions is excellent.

\medskip

Measurements of the various neutrino-beam characteristics
at the KEK site agree with the Monte-Carlo expectations,
which ensures that the comparison
of neutrino events in SK with expectations
based on the same Monte-Carlo simulation is relaiable.

\begin{figure}[b!]
\centerline{
\psfig{file=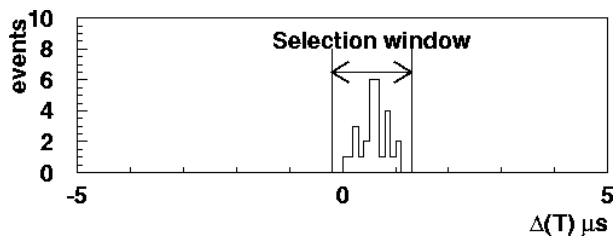,height=3.cm}
}
\caption{
Time correlation between the neutrino beam period and SK events
which are selected by the standard atmospheric neutrino analysis.
Events in the 1.5~$\mu$sec gate are finally selected.
}
\label{fig4}
\end{figure}

\section{Observation in Super-Kamiokande}

To obtain beam-correlated fully contained neutrino interactions,
an event selection similar to an atmospheric neutrino
analysis\cite{SuperK} was applied; 
the time correlation with the neutrino beam was then examined.
Figure~\ref{fig4} shows the time difference between the neutrino beam
and the events obtained from atmospheric neutrino selection.

Considering the neutrino beam duration (1.1$\mu$sec)
and accuracy\cite{UTC} of
the absolute time determination ($<0.2\mu$sec), events within a 1.5$\mu$sec
time window covering the neutrino beam period were selected.
A total of 28 fully-contained events were found
in 22.5kt of the fiducial volume.
Details of the neutrino events are summarized in Table.1.
Because the expected atmospheric neutrino background in the fiducial
volume within the neutrino beam period was calculated to
be $6\times 10^{-4}$ events,
the 28 events in the fiducial volume are a clear signal of neutrinos
from KEK.

\begin{table}[b!]
\caption{
Number of neutrino events in SK. Expectations based on the
event rate at 1KT, SFT, and MRD are also shown
}
\begin{center}
\begin{tabular}{lrrrr}
\hline
\hline
Event Category~~~~~~~~~~~~& SK data~~~~~~ && Expected & \\
                          &         &(1KT)~~~~&(SFT)~~~~&(MRD)~~~~\\
\hline
Single ring events        & 15~~~~~~~~~& 22.9~~~~~~~&& \\
~~(e-like)                & 14~~~~~~   & 20.9~~~~   && \\
~~($\mu$-like)            &  1~~~~~~   &  2.0~~~~   && \\
Multi ring events         & 13~~~~~~~~~& 14.9~~~~~~~&& \\
\hline
Total                        & 28~~~~~~~~~& 37.8\hbox{${{+3.5}\atop{-3.8}}$}
 & 37.2\hbox{${{+4.6}\atop{-5.0}}$} & 41.0\hbox{${{+6.0}\atop{-6.6}}$} \\
\hline
\hline
\end{tabular}
\end{center}
\end{table}

\section{Oscillation analysis}

Strategies concerning oscillation searches at K2K are summarized as follows. 
The \nmnt oscillation can be examined by a disappearance of neutrino events
in SK, because the energy of the neutrino beam is smaller
than the $\tau$ production threshold.
In addition, the neutrino energy spectrum in SK should be
distorted in the case of oscillation, because the oscillation probability
is a function of the neutrino energy.

An examination of the \nenm oscillation is an appearance search.
A possible excess of $\nue$ events in SK is
direct evidence of the \nenm oscillation, because the original beam from KEK
is almost pure $\num$, and because
the particle identification capability in SK
is excellent.\cite{SuperK,Kasuga}
In addition, the total neutrino interactions should be almost equal
to the expectation with a null oscillation, because $\nue$
also interacts through charged-current interactions.

The following three subsections discuss the present status of the data
analyses about three subjects, i.e. 
(1)absolute event number, (2)$\nue/\num$ ratio,
and (3)distortion of the neutrino energy spectrum.

\subsection{Absolute event numbers}

The expected event numbers in SK can be calculated
from the neutrino event rate in FDs, and an extrapolation from the FDs to SK.
The numbers obtained in 1KT, SFT, and MRD are used
for the event rate in the FDs, as reported in ${\it 2.2}$. 
The extrapolation is calculated from the neutrino beam intensity and
its angular divergence obtained from the pion monitor.
The expectations based on the data from the FDs
are 37.8\hbox{${{+3.5}\atop{-3.8}}$}(1KT),
37.2\hbox{${{+4.6}\atop{-5.0}}$}(SFT), and
41.0\hbox{${{+6.0}\atop{-6.6}}$}(MRD).
These results are consistent with each other.
We used the numbers from 1KT as an official number
because of its small systematic errors.  

The statistical probability that the observation is equal
to, or smaller than, 28, where the expectation is
37.8\hbox{${{+3.5}\atop{-3.8}}$}, is 9.6\%.
The observation is slightly smaller than the expectation,
but it is not statistically significant.

\begin{figure}[b!]
\centerline{
\epsfig{file=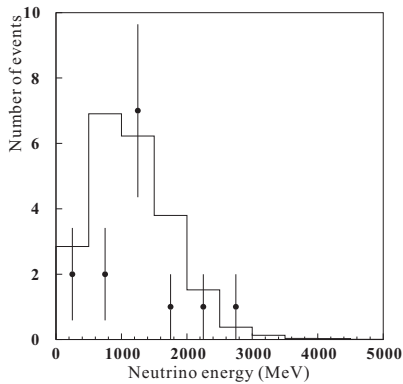,height=5.cm}
}
\caption{
Neutrino energy distribution for 14 single ring $\mu$-like events in SK.
Monte Carlo expectation for no oscillation is also shown.
}
\label{fig5}
\end{figure}

\subsection{$\nue/\num$ ratio}

The single ring events in SK are judged $\mu$-like or e-like
by the standard particle-identification program\cite{Kasuga}
developped for the SK atmospheric neutrino analysis.
As shown in Table~1, the number of $\mu$-like and e-like events are
14 and 1, respectively.
Since the expected e-like events was calculated to be 2.0, there is
no excess in the number of e-like events.
The reduction of the total event numbers as well as the agreement of
the $\nue$ events indicate that a pure \nenm oscillation is not the solution.
However, more statistics is needed to give a conclusion.
The possibilities of the 3 flavor oscillation ($\nue$, $\num$ and $\nut$)
must also be studied.\cite{Yoshida}

\subsection{Distortion of neutrino energy spectrum}

Neutrino energy spectrum calculated from 14 single ring $\mu$-like events are
shown in Figure~5 together with an expectation obtained from
the pion monitor data and a Monte-Carlo simulation.
At present, the statistics is too poor to examine
the neutrino energy spectrum.   

\section{Summary}

The K2K long-baseline neutrino-oscillation experiment
has been successfully operated since 1999.
By the end of 2000, a total intensity of 22.9$\times 10^{18}$
protons on target were accumulated, which is about 20\% of the
goal of the experiments. A total of 28 fully-contained neutrino
interactions in the 22.5kt of the fiducial volume of the
Super-Kamiokande detector were observed,
where the expectation based on the data from the Front Detectors
is 37.8\hbox{${{+3.5}\atop{-3.8}}$}.
The statistical probability that the observation is equal
to or smaller than 28 for the expectation of
37.8\hbox{${{+3.5}\atop{-3.8}}$} is 9.6\%.
Although the observation is slightly smaller than the expectation
and it is faint evidence of neutrino oscillations,
the descrepancy is still within the statistical error,
and is not significant.
Oscillation analyses based on $\nue/\num$ ratio and distortion of the
neutrino energy spectrum are also in progress.

\end{document}